\documentclass[12pt]{article}
\usepackage[cp1251]{inputenc}
\usepackage{amsmath}
\usepackage{amsfonts}
\usepackage{amssymb}
\def\pa{\partial}                       
\def\beq{\begin{eqnarray}}    
\def\eeq{\end{eqnarray}}      

\begin{document}
\date{}

\begin{center}
{\Large\textbf{On classical and quantum deformations of gauge theories}}

\vspace{18mm}

{\large I.L. Buchbinder$^{(a,b)}\footnote{E-mail:
joseph@tspu.edu.ru}$\;,
P.M. Lavrov$^{(a,b,)} \footnote{E-mail:
lavrov@tspu.edu.ru}$,\;
}

\vspace{8mm}

\noindent  ${{}^{(a)}}
${\em
Center of Theoretical Physics, \\
Tomsk State Pedagogical University,\\
Kievskaya St.\ 60, 634061 Tomsk, Russia}

\noindent  ${{}^{(b)}}
${\em
National Research Tomsk State  University,\\
Lenin Av.\ 36, 634050 Tomsk, Russia}

\vspace{20mm}

\begin{abstract}
\noindent
We elaborate the generalizations of the approach to
gauge-invariant deformations of the gauge theories developed in our
previous work \cite{BL-21}. In the given paper we construct the
exact transformations defying the gauge-invariant deformed theory on
the base of initial gauge theory with irreducible open gauge
algebra. Like in \cite{BL-21}, for the theories with open gauge
algebras these transformations are the shifts of the initial gauge
fields $A \rightarrow A+h(A)$, with the help of the arbitrary and in
general non-local functions $h(A)$. The results are applied to study
the quantum aspects of the deformed theories. We derive the exact
relation between the quantum effective actions for the above
classical theories, where one is obtained from another with the help of
the deformation.

\end{abstract}

\end{center}

\vfill

\\

\newpage

\section{Introduction}
\noindent
The gauge theories play a key role in modern
theoretical and mathematical physics. Therefore, the study of new
classical and quantum aspects of gauge theories still deserves a
close attention. One of such new aspects has been considered in our
recent paper \cite{BL-21}, where a gauge-invariant
deformation of classical gauge theories was worked out. To be more precise, we have
described a deformation of the classical gauge theory with a closed
gauge algebra, which again leads to a gauge theory with a closed
algebra. It was pointed out in \cite{BL-21} that this deformation
can be local or non-local. In the first case the deformation results in a
theory equivalent to the initial one, in the second case, we obtain a new
gauge theory with new gauge algebra. In some case such a non-local
theory posses a closed local sector, forming a new local gauge
theory. As a result, the deformation under consideration can be
treated as method to generate the new gauge theories based
given gauge theories. In particular, it was shown in \cite{BL-21}
that the Yang-Mills theory is obtained as the deformation of abelain
gauge theory, and a cubic interaction vertex in bosonic higher spin
theory is generated as the deformation of free bosonic higher spin
theory.

Derivation of the deformation in \cite{BL-21} was based on the
Batalin-Vilkovisky (BV) formalism \cite{BV}, \cite{BV1}, \cite{BV2},
where the central object is the master equation, allowing to explore
the classical and quantum aspects of gauge theories by some
universal way\footnote{See the recent developments and applications
of the BV-formalism in \cite{BLT-15}, \cite{BL-16}, \cite{BLT-21}
and the references therein.}. Construction of the gauge-invariant
deformation is closely related with the study of the classical
master equation. The solution to this equation was investigated in
the works \cite{BH}, \cite{H} in the form of an expansion over some
parameter and was shown that finding the solution reduces to
cohomology problems. Applications of such an approach to Yang-Mills
theory and higher spin field theory were considered in the works
\cite{D}, \cite{BaBu} and \cite{BL},\cite{BLS}, \cite{SS}
respectively. Unlike all these works, our approach does not require
to find a solution to the master equation in the form of series but
allows to construct a general solution to master equation in closed
form.

In this paper, we continue to explore the gauge-invariant deformation
of the gauge theories and study the two new aspects. First, we
generalize the results of the work \cite{BL-21} for gauge theories
with irreducible generators and open gauge algebra. The examples of
such theories are the supergravity models in component formulation,
where the local superalgebras are only on-shell closed\footnote{For
some supergravity models there exists the superfield formulations
where the local supersymmetry transformations form a closed algebra
(see e.g. \cite{Super}, \cite{BK}, \cite{Harmonic}). However for
arbitrary supergravity models a formulation in terms of
unconstrained superfields is still unknown.}. Second, using the
above results for classical deformations, we study the quantum
aspects of the deformed theories and derive the exact relation
between quantum effective actions of initial and deformed gauge
theories.

The paper is organized as follows. In section 2, we consider the
basic notions of classical theory with irreducible open gauge
algebras within the BV-formalism. Section 3 is devoted to deriving
the possible deformations of the initial classical gauge theory with
the help of minimal anticanonical transformations. In section 4, we
consider the quantization of the initial classical gauge theory and
the gauge-invariant deformed theories using the BV-procedure and
find the relationship between the effective action in these
theories. In section 5, we summarize the results.

In the paper we systematically use the DeWitt's condensed notations
and employ the symbols $\varepsilon(A)$ for the Grassmann parity and
${\rm gh}(A)$ for the ghost number, respectively. The right and left
functional derivatives are marked by special symbols $"\leftarrow"$
and $"\overrightarrow{}"$ respectively. Arguments of any functional
are enclosed in square brackets $[\;]$, and arguments of any
function are enclosed in parentheses $(\;)$.

\section{Classical gauge theory in the BV- formalism }
\noindent We consider a classical gauge theory of the fields
$A=\{A^i\}$ with the Grassmann parities
$\varepsilon(A^i)=\varepsilon_i$ and the ghost numbers $\;{\rm
gh}(A^i)=0$. The theory is described by an initial action $S_0[A]$
invariant under the gauge transformations $\delta
A^i=R^i_{\alpha}(A)\xi^{\alpha}$, \beq \label{S0giv}
S_{0,i}[A]R^i_{\alpha}(A)=0,\quad
S_{0,i}[A]=S_{0}[A]\overleftarrow{\pa}_{\!\!A^i}, \eeq where
$R^i{}_{\alpha}(A)$
($\varepsilon(R^i_{\alpha}(A))=\varepsilon_i+\varepsilon_{\alpha},\;
{\rm gh}(R^i_{\alpha}(A))=0$) are the gauge generators and the gauge
parameters $\xi^{\alpha}$
($\varepsilon(\xi^{\alpha})=\varepsilon_{\alpha}$) are the arbitrary
functions of space-time coordinates. It is assumed that the gauge
generators obey an irreducible and, generally speaking, open
gauge algebra,
\beq
\label{ga}
R^i_{\alpha ,
j}(A)R^j_{\beta}(A)-(-1)^{\varepsilon_{\alpha}\varepsilon_
{\beta}}R^i_{\beta ,j}(A)R^j_{\alpha}(A)=-R^i_{\gamma}(A)
F^{\gamma}_{\alpha\beta}(A)-S_{0,j}M^{ji}_{\alpha\beta}(A),
\eeq
where $R^i_{\alpha ,
j}(A)=R^i_{\alpha}(A)\overleftarrow{\pa}_{\!\!A^j}$,
$F^{\gamma}_{\alpha\beta}(A)$
($\varepsilon(F^{\gamma}_{\alpha\beta}(A))=\varepsilon_{\alpha}+
\varepsilon_{\beta}+\varepsilon_{\gamma},\;{\rm
gh}(F^{\gamma}_{\alpha\beta}(A))=0$) are the structure coefficients
depending, in general, on the fields $A^i$, with the following
symmetry properties $F^{\gamma}_{\alpha\beta}(A)=
-(-1)^{{\varepsilon_{\alpha}\varepsilon_{\beta}}}F^{\gamma}_{\beta\alpha}
(A)$. The functions $M^{ij}_{\alpha\beta}(A)$ satisfy the conditions
$M^{ij}_{\alpha\beta}(A) = -(-1)^{\varepsilon_i\varepsilon_j}
M^{ji}_{\alpha\beta}(A) =
-(-1)^{\varepsilon_{\alpha}\varepsilon_{\beta}}M^{ij}_{\beta\alpha}(A)$.
If $M^{ij}_{\alpha\beta}(A)\neq
0$, then the gauge algebra is called open. Otherwise, the algebra is called closed. If
$M^{ij}_{\alpha\beta}(A) = 0$, and $F^{\gamma}_{\alpha\beta}$ does
not depend on the fields, the gauge transformations form a gauge
group  and define a Lie algebra.

Following the BV-formalism \cite{BV,BV1}, we introduce the minimal
antisymplectic space of fields $\phi^A_m=(A^i,C^{\alpha})$ and
antifields $\phi^*_{m\;\!A}=(A^*_i,C^*_{\alpha}).$ Here $C^{\alpha}$
($\varepsilon(C^{\alpha})=\varepsilon_{\alpha}+1,\; {\rm
gh}(C^{\alpha})=1$) are the ghost fields. The antifields
$\phi^*_{m\;\!A}$ obey the following properties
\beq
\varepsilon(\phi^*_{m\;\!A})=\varepsilon(\phi^A)+1, \quad {\rm
gh}(\phi^*_{m\;\!A})=-1-{\rm gh}(\phi^A).
\eeq
The basic object of
the BV-formalism is the extended action $S_m=S_m[\phi_m,\phi^*_m]$
satisfying the classical master equation, \footnote{For any set of
fields $\phi^A$ and antifields $\phi^*_A$ and arbitrary functionals
$G=G[\phi,\phi^*]$ and $H=H[\phi,\phi^*]$ the antibracket is defined
by the rule $(G,H)=
G\left(\overleftarrow{\pa}_{\!\!\phi^A}\overrightarrow{\pa}_{\!\!\phi^*_{A}}-
\overleftarrow{\pa}_{\!\!\phi^*_{A}}\overrightarrow{\pa}_{\!\!\phi^A}\right)H.$}
\beq
\label{master}
(S_m,S_m)=0,
\eeq
and the boundary condition,
\beq
\label{boundary} S_m[\phi_m,\phi^*_m]\Big|_{\phi^*_m=0}=S_0[A].
\eeq
The action $S_m$ up to the terms linear in antifields is
written in the form
\beq
\label{S}
S_m=S_0[A]+A^*_iR^i_{\alpha}(A)C^{\alpha}-
\frac{1}{2}C^*_{\gamma}F^{\gamma}_{\alpha\beta}(A)C^{\beta}C^{\alpha}
(-1)^{\varepsilon_{\alpha}}+O(\phi^{*\;2}).
\eeq
It is proved that
the solutions to the classical master equation in the minimal
antisymplectic space encode all structure functions of the gauge
algebra of a given classical gauge theory \cite{BV,BV1}.

The next step of the BV-formalism \cite{BV,BV1} involves introduction
of the full antisymplectic
space of fields  $\phi^A$ and antifields  $\phi^*_A$. For irreducible gauge theories
the set of fields  and antifields  reads
\beq
\label{fantisymp}
\phi^A=(A^i,C^{\alpha}, {\bar C}^{\alpha}, B^{\alpha}), \quad
\phi^*_{A}=(A^*_i,C^*_{\alpha}, {\bar C}^*_{\alpha}, B^*_{\alpha}),
\eeq
where ${\bar C}^{\alpha}$ $(\varepsilon({\bar C}^{\alpha})=1,
{\rm gh}({\bar C}^{\alpha})=-1)$ are the anti-ghost fields and $B^{\alpha}$
$(\varepsilon({B}^{\alpha})=0,
{\rm gh}({B}^{\alpha})=0)$ are the auxiliary (Nakanishi-Lautrup) fields.
In the space of the fields (\ref{fantisymp}) the action $S=S[\phi,\phi^*]$
\beq
\label{fsolact}
S=S_m+{\bar C}^*_{\alpha}B^{\alpha}
\eeq
is introduced. This action satisfies the classical master equation
\beq
(S,S)=0 \quad {\rm with}\quad S[\phi,\phi^*]\big|_{\phi^*=0} = S_0[A].
\eeq
The gauge invariance of the initial action $S_0[A]$ leads to
invariance of the action $S=S[\phi,\phi^*]$,
\beq
\delta_B S=0
\eeq
under the BRST transformations
\cite{brs1,t}
\beq
\delta_B \phi^A=(\phi^A,S)\mu=
\overrightarrow{\pa}_{\!\!\phi^*_{A}}S\;\mu,\quad \delta_B
\phi^*_{A}=0.
\eeq
As a consequence of these relations, the $S$ satisfies the classical
master equation. Here $\mu$ is a constant Grassmann parameter.

It is proved that there exist the specific transformations of fields
and antifields, preserving the antibracket, which are called
anticanonical \cite{BV,BV1}. It means that making an arbitrary
anticanonical transformation in solution to the classical master
equation, we obtain an action satisfying again to the classical
master equation. Namely, this fact was used to construct a
gauge-invariant deformation of the gauge theories \cite{BL-21}.

\section{Gauge-invariant deformation of classical theory}
\noindent
Gauge invariant deformation of the classical gauge theory was stated in our paper
\cite{BL-21} for the theory with closed gauge algebra. Now we generalize the results of \cite{BL-21}
for the theory where gauge algebra is open.

According to the results of the paper \cite{BL-21}, the deformation
of initial theory in the full antisymplectic space is described by
the following anticanonical transformations of the solution
(\ref{fsolact}),
\beq
\label{transformation}
\phi^*_{A}=Y[\phi,\Phi^*]\overleftarrow{\pa}_{\!\!\phi^A},\quad
\Phi^A=\overrightarrow{\pa}_{\!\!\Phi^*_A}Y[\phi,\Phi^*], \eeq where
$Y=Y[\phi,\Phi^*]$ ($\varepsilon(Y)=1,\;{\rm gh}(Y)=-1$) is the
generating functional of the anticanonical transformation.
These transformations are non-trivial in the sector
of minimal antisymplectic space only and, in general, are described
by the generating functional of the form\footnote{Note that the form of transformations
(\ref{antiCT}) is the same both for closed and for open
gauge algebras.}
\beq
\label{antiCT}
Y[\phi,\Phi^*]=\Phi^*_A\phi^A+ {\cal A}^*_ih^i(A)+ {\cal
C}^*_{\alpha}g^{\alpha}_{\;\beta}(A)C^{\beta},
\eeq
where $h^i(A)$
$(\varepsilon(h^i(A))=\varepsilon_i, {\rm gh}(h^i(A))=0)$ and
$g^{\alpha}_{\;\beta}(A)$
$(\varepsilon(g^{\alpha}_{\;\beta}(A))=\varepsilon_{\alpha}+\varepsilon_{\beta},
{\rm gh}(g^{\alpha}_{\;\beta}(A))=0)$ are the arbitrary functions of
fields $A^i$. Here the following notations
\beq
\Phi^A=({\cal A}^i,
{\cal C}^{\alpha}, {\bar{\cal C}}^{\alpha}, {\cal B}^{\alpha}),\quad
\Phi^*_A=({\cal A}^*_i,{\cal C}^*_{\alpha}, {\bar{\cal
C}}^*_{\alpha}, {\cal B}^*_{\alpha})
\eeq
are used.

It was proved in \cite{BL-21} that the anticanonical transformations (\ref{antiCT})
of the initial classical action $S_0[A]$ lead to the deformed
action $\widetilde{S}_0[A]$ of the form
\beq
\label{deformaction}
\widetilde{S}_0[A]=S_0[A+h(A)].
\eeq
It is worth emphasizing that the deformed action (\ref{deformaction})
has extremely simple form and
is described by the functions $h^i(A)$ only. The action $\widetilde{S}_0[A]$
is invariant under gauge transformations,
$\delta A^i=\widetilde{R}^i_{\alpha}(A)\xi^{\alpha}$,
\beq
\widetilde{S}_{0,i}[A]\widetilde{R}^i_{\alpha}(A)=0.
\eeq
Here
\beq
\label{gdGen}
\widetilde{R}^i_{\alpha}(A)=(M^{-1}(A))^i_{\;j}R^j_{\alpha}(A+h(A))
\eeq
are deformed gauge generators and the $(M^{-1}(A))^i_{\;j}$ is inverse to the matrix
\beq
\label{M}
M^i_{\;j}(A)=\delta^i_{\;j}+h^i(A)\overleftarrow{\pa}_{\!A^j} .
\eeq
The deformed gauge
generators (\ref{gdGen}) are formulated in terms of functions $h^i(A)$ only. The
functions $g^{\alpha}_{\;\beta}(A)$ are responsible for possible transformations of ghost
fields $C^{\alpha}$ and antifields $C^*_{\alpha},$ which do not touch the deformations
in the sector of fields $A^i$. It allows us to introduce the {\it minimal} anticanonical
transformations describing by the generating functional
\beq
\label{mantiCT}
Y_{min}[\phi,\Phi^*]=\Phi^*_A\phi^A+ {\cal A}^*_ih^i(A).
\eeq
In this case the procedure of gauge-invariant deformations of initial gauge-invariant
theory is controlled by the functions $h^i(A)$ only. It means that further there is no need to use
the functions $g^{\alpha}_{\;\beta}(A)$.
As a result, the gauge-invariant deformations corresponds to the following
non-trivial change of variables
\beq
A^i\quad \rightarrow \quad A^i+h^i(A), \qquad
A^*_i \quad \rightarrow \quad A^*_j (M^{-1}(A))^j_{\;\;\!i}\;.
\eeq

The gauge algebra of deformed generators (\ref{gdGen}) can be
calculated the same way as it was done in the paper \cite
{BL-21}, but taking into account that the initial gauge algebra is
now open. The result has the form
\beq
\label{dga}
\widetilde{R}^i_{\alpha ,
j}(A)\widetilde{R}^j_{\beta}(A)-(-1)^{\varepsilon_{\alpha}\varepsilon_
{\beta}}\widetilde{R}^i_{\beta ,j}(A)\widetilde{R}^j_{\alpha}(A)=
-\widetilde{R}^i_{\gamma}(A)
\widetilde{F}^{\gamma}_{\alpha\beta}(A)-\widetilde{S}_{0,j}[A]
\widetilde{M}^{ji}_{\alpha\beta}(A),
\eeq
Here $\widetilde{F}^{\gamma}_{\alpha\beta}(A)=F^{\gamma}_{\alpha\beta}(A+h(A))$
are the deformed structure functions and the functions $\widetilde{M}^{ji}_{\alpha\beta}(A)$ are
\beq
\widetilde{M}^{ji}_{\alpha\beta}(A)=
(M^{-1}(A))^j_{\;l}(M^{-1}(A))^i_{\;k}M^{kl}_{\alpha\beta}(A+h(A))
(-1)^{\varepsilon_l(\varepsilon_i+\varepsilon_k)}.
\eeq
The symmetry properties of deformed structures
$\widetilde{F}^{\gamma}_{\alpha\beta}(A)$ and
$\widetilde{M}^{ji}_{\alpha\beta}(A)$ coincide with corresponding
ones of non-deformed structures. The same is valid for the
Grassmann parities and ghost numbers. We see that the deformed
algebra is again irreducible and open.

After the anticanonical transformations (\ref{mantiCT}), the action
(\ref{fsolact}) transforms into the functional
$\widetilde{S}=\widetilde{S}[\phi,\phi^*]$ which satisfies the
classical master equation \beq \label{dactS}
(\widetilde{S},\widetilde{S})=0 \eeq and has the following form up
to the terms linear in antifields \beq \label{dfact}
\widetilde{S}=\widetilde{S}_0[A]+A^*_i\widetilde{R}^i_{\alpha}(A)C^{\alpha}-
\frac{1}{2}C^*_{\gamma}\widetilde{F}^{\gamma}_{\alpha\beta}(A)C^{\beta}C^{\alpha}
(-1)^{\varepsilon_{\alpha}}+{\bar
C}^*_{\alpha}B^{\alpha}+O(\phi^{*\;2}). \eeq The action
(\ref{dfact}) is invariant under the BRST transformations, \beq
\delta_B \widetilde{S}=0,\quad \delta_B
\phi^A=(\phi^A,\widetilde{S})\mu=
\overrightarrow{\pa}_{\!\!\phi^*_{A}}\widetilde{S}\;\mu,\quad
\delta_B \phi^*_{A}=0, \eeq due to the equation (\ref{dactS}). As a
result, we have described the gauge-invariant deformation of
classical gauge theory for the case of open gauge algebra. If the
initial gauge algebra is closed, the result (\ref{dga}) is reduced
to deformed gauge algebra constructed in the paper \cite {BL-21} for
the special case $g^{\alpha}_{\;\beta}(A)=0$.

\section{Deformation of quantum gauge theory}
\noindent
This section derives the general relationship between the quantum
effective actions of two
classical gauge theories obtained from each other using the deformation procedure
described in the previous section. To be more precise, we will consider
the effective actions of two such theories in the sector of fields $A^i$ only.

According to BV-formalism \cite{BV,BV1}, the generating functional
of Green functions in the case under consideration is given by the
following functional integral
\beq
\label{iZ}
Z[j]=\int D\phi
\exp\Big\{\frac{i}{\hbar}\big(S_g[\phi]+jA\big)\Big\},
\eeq
where $D\phi=DADCD{\bar C}DB$ and
\beq
\label{Sg}
S_g[\phi]=S[\phi,\phi^*]|_{\phi^*=\psi[\phi]\overleftarrow{\pa_{\phi}}}
\eeq
is the gauge fixed action. Here $j=\{j_i\}$
$(\varepsilon(j_i)=\varepsilon_i, {\rm gh}(j_i)=0)$ is the set of
external sources to the fields $A^i$, $\psi[\phi]$
($\varepsilon(\psi[\phi])=1, {\rm gh}(\psi[\phi])=1)$ is the gauge
fixing functional which is chosen standardly as $\psi[\phi]={\bar
C}^{\alpha}\chi_{\alpha}(A),$ where $\chi_{\alpha}(A)$
$(\varepsilon(\chi_{\alpha}(A))=\varepsilon_{\alpha}, {\rm
gh}(\chi_{\alpha}(A))=0)$ are the gauge fixing functions  lifting
the degeneracy of action $S[\phi,\phi^*]$. It allows to write
\beq
\label{actg}
S_g[\phi]=S_0[A]+{\bar
C}^{\alpha}\chi_{\alpha,i}(A)R^i_{\beta}(A)C^{\beta}+
\chi_{\alpha}(A)B^{\alpha}+O(C^2).
\eeq
For Yang-Mills type gauge
theories the first three terms in the r.h.s of (\ref{actg}) present
the well known Faddeev-Popov action \cite{FP}.

According to the same procedure, the generating functional of Green
functions  of the deformed theory has the form
\beq
\label{dZ}
\widetilde{Z}[j]=\int D\phi
\exp\Big\{\frac{i}{\hbar}\big(\widetilde{S}_g[\phi]+jA\big)\Big\},
\eeq
where
\beq
\widetilde{S}_g[\phi]=\widetilde{S}[\phi,\phi^*]|_{\phi^*=
\widetilde{\psi}[\phi]\overleftarrow{\pa_{\phi}}}
\eeq
is the gauge
fixed action for deformed theory corresponding to the gauge fixing
condition subjected to the change of fields $A^i$ coming from the
minimal anticanonical transformations
\beq
\widetilde{\psi}[\phi]=\psi[A+h(A), {\bar C}].
\eeq

In this case, we have the simple relation
\beq
\label{dSiS}
\widetilde{S}_g[\phi]=S_g[\phi]\big|_{A\rightarrow A+h(A)}
\eeq
and,
therefore,
\beq
\label{dactg}
\widetilde{S}_g[\phi]=\widetilde{S}_0[A]+ {\bar
C}^{\alpha}\widetilde{\chi}_{\alpha,i}(A)\widetilde{R}^i_{\beta}(A)C^{\beta}+
\widetilde{\chi}_{\alpha}(A)B^{\alpha}+O(C^2).
\eeq

Let us make in
the functional integral (\ref{iZ}) the change of variables
\beq
\label{trA}
A^i\quad\rightarrow \quad A^i+h^i(A),
\eeq
and take into
account the relation (\ref{dSiS}) as well as the Jacobian $J[A]$
corresponding to the transformation (\ref{trA})
\beq
\label{jacobian} J[A]={\rm sDet} M(A)=\exp\{{\rm sTr}\ln M(A)\},
\eeq
where the symbols ${\rm sDet}$ and ${\rm sTr}$ mean the
functional superdeterminant and functional supertrace respectively.
After that we arrive at the relation
\beq
\label{iZdZ}
Z[j]=\;:\exp\Big\{\frac{i}{\hbar}jh(\widehat{q})\Big\}:
\exp\big\{{\rm sTr}\ln M(\widehat{q})\big\}\widetilde{Z}[j].
\eeq
T

The relation (\ref{iZdZ}) allows to express the Green functions of the initial
theory through the Green functions in the deformed theory. In the
equation (\ref{iZdZ}) the matrix $M(A)$ is defined in (\ref{M}) and
the notation
\beq
\widehat{q}^i=\frac{\hbar}{i}\frac{\overrightarrow{\delta}}{\delta
j_i}
\eeq
is used. The symbol $:\;:$ means
\beq
\label{::}
:\exp\{j\widehat{q}\}:= \sum_{n=0}^{\infty}\frac{1}{n!}j_1j_2\cdots
j_n\widehat{q}_n\cdots \widehat{q}_2\widehat{q}_1 ,
\eeq
i.e all operators $\widehat{q}^i$ must stand from the right of sources $j_i$
in the order indicated exactly in (\ref{::}).

Also we can find the relation inverse to (\ref{iZdZ}). To do this let us act by the
operator $:\exp\{(i/\hbar)jh(\widehat{q})\}:$ on (\ref{dZ}) from the
left side. It leads to
\beq
\label{::dZ}
:\exp\Big\{\frac{i}{\hbar}jh(\widehat{q})\Big\}:
\widetilde{Z}[j]=\int D\phi
\exp\Big\{\frac{i}{\hbar}\big(\widetilde{S}_g[\phi]+jA+jh(A)\big)\Big\}.
\eeq
Making use in the functional integral (\ref{::dZ}) the change
of variables
\beq
A^i+h^i(A) \quad \rightarrow \quad A^i ,
\eeq
we obtain
\beq
:\exp\{\frac{i}{\hbar}jh(\widehat{q})\}:\widetilde{Z}[j]= \exp
\{-{\rm sTr}\ln M(\widehat{q})\}Z[j].
\eeq
Finally, the relation inverse to (\ref{iZdZ}) reads
\beq
\label{dZiZ}
\widetilde{Z}[j]=:\exp\{-\frac{i}{\hbar}jh(\widehat{q})\}: \exp
\{-{\rm sTr}\ln M(\widehat{q})\}Z[j] .
\eeq

Next step of consideration is to introduce the generating functionals of the connected Green functions
\beq
W[j]=-i\hbar \ln Z[j], \quad \widetilde{W}[j]=-i\hbar \ln \widetilde{Z}[j].
\eeq
From the equation (\ref{iZdZ})  it follows
\beq
\label{iWdW}
W[j]=\widetilde{W}[j]-i\hbar
\ln\Big(:\exp\Big\{\frac{i}{\hbar}jh(\widehat{\widetilde{Q}})\Big\}:
\exp\{{\rm sTr}\ln M(\widehat{\widetilde{Q}})\}\Big) \cdot 1.
\eeq
where the operators $\widehat{\widetilde{Q}^i}$ are
\beq
\widehat{\widetilde{Q}^i}=
\exp\Big\{-\frac{i}{\hbar}\widetilde{W}[j]\Big\}\widehat{q}^i
\exp\Big\{\frac{i}{\hbar}\widetilde{W}[j]\Big\}=
\widehat{q}^i+\frac{\overrightarrow{\delta}\widetilde{W}[j]}{\delta j_i}.
\eeq
Here the following commutation relations
\beq
[q^i,q^j]=0 \quad \rightarrow \quad [\widehat{Q}^i,\widehat{Q}^j]=0
\eeq
were used in deriving the equation (\ref{iWdW}).
In its turn the equation (\ref{dZiZ}) in terms of functionals $W[j]$
and $\widetilde{W}[j]$ rewrites in the form
\beq
\label{dWiW}
\widetilde{W}[j]=W[j]-i\hbar
\ln\Big(:\exp\Big\{-\frac{i}{\hbar}jh(\widehat{Q})\Big\}:
\exp\{-{\rm sTr}\ln M(\widehat{Q})\}\Big) \cdot 1,
\eeq
where
\beq
\widehat{Q}^i=
\exp\Big\{-\frac{i}{\hbar}W[j]\Big\}\widehat{q}^i
\exp\Big\{\frac{i}{\hbar}W[j]\Big\}=
\widehat{q}^i+\frac{\overrightarrow{\delta}W[j]}{\delta j_i}.
\eeq
One can check that the relation $[\widehat{Q}^i, \widehat{Q}^j]=0$ is fulfilled as it should be.

Now one introduces the effective actions using the
Legendre transformations of functionals $W[j]$ and $\widetilde{W}[j]$,
\beq
\label{G}
&&\Gamma[{\cal A}]=W[j]-j{\cal A}, \quad {\cal A}^i=
\frac{\overrightarrow{\delta}W[j]}{\delta j_i},\quad
\Gamma[{\cal A}]\overleftarrow{\pa}_{\!{\cal A}^i}=-j_i,\\
&&
\label{dG}
\widetilde{\Gamma}[\widetilde{{\cal A}}]=\widetilde{W}[j]-j\widetilde{{\cal A}},
\quad \widetilde{{\cal A}}^i=
\frac{\overrightarrow{\delta}\widetilde{W}[j]}{\delta j_i},\quad
\widetilde{\Gamma}[\widetilde{{\cal A}}]\overleftarrow{\pa}_{\!\widetilde{{\cal A}}^i}=-j_i.
\eeq
Then, the relation (\ref{iWdW}) leads to the expressions
\beq
\nonumber
&&\Gamma[{\cal A}]-\Gamma[{\cal A}]\overleftarrow{\pa}_{\!{\cal A}^i}{\cal A}^i=
\widetilde{\Gamma}[\widetilde{{\cal A}}]-
\widetilde{\Gamma}[\widetilde{{\cal A}}]\overleftarrow{\pa}_{\!\widetilde{{\cal A}}^i}
\widetilde{{\cal A}}^i-\\
\label{iGdG}
&&
-i\hbar \ln\big(:\exp\{-\frac{i}{\hbar}(\widetilde{\Gamma}[\widetilde{{\cal A}}]
\overleftarrow{\pa}_{\!\widetilde{{\cal A}}})h(\widehat{\widetilde{{\cal A}}})\}:
\exp\{{\rm sTr}\ln M(\widehat{\widetilde{{\cal A}}})\}\big) \cdot 1,
\eeq
where the notation
\beq
\widehat{\widetilde{{\cal A}^i}}=\widetilde{{\cal A}}^i+i\hbar
(\widetilde{\Gamma}^{(-1)}[\widetilde{{\cal A}}])^{ij}
\frac{\overrightarrow{\delta}}{\delta \widetilde{A}^j}
\eeq
is used. The matrix $(\widetilde{\Gamma}^{(-1)}[\widetilde{{\cal A}}])^{ij}$
is inverse to the matrix
\beq
(\widetilde{\Gamma}[\widetilde{\cal A}])_{ij}=
\frac{\overrightarrow{\delta}}{\delta {\cal A}^i}
\widetilde{\Gamma}[\widetilde{A}]\frac{\overleftarrow{\delta}}{\delta {\cal A}^j},\qquad
(\widetilde{\Gamma}[\widetilde{\cal A}])_{ik}
(\widetilde{\Gamma}^{(-1)}[\widetilde{{\cal A}}])^{kj}=\delta_i^{\;j},
\eeq
with the following symmetry properties
\beq
\label{sym}
(\widetilde{\Gamma}^{(-1)}[\widetilde{{\cal A}}])^{ij}=-
(\widetilde{\Gamma}^{(-1)}[\widetilde{{\cal A}}])^{j\;\!i}
(-1)^{(\varepsilon_i+1)(\varepsilon_j+1)},\quad
(\widetilde{\Gamma}[\widetilde{\cal A}])_{ij}=-
(\widetilde{\Gamma}[\widetilde{\cal A}])_{j\;\!i}
(-1)^{(\varepsilon_i+1)(\varepsilon_j+1)}.
\eeq
Note that the operators $\widehat{\widetilde{{\cal A}^i}}$ commute,
\beq
[\widehat{\widetilde{{\cal A}^i}}, \widehat{\widetilde{{\cal A}^j}}]=0,
\eeq
due to the relations (\ref{sym}).
Differentiating the relation (\ref{iWdW}) with respect to sources $j_i$  and
taking into account the rule for derivative of the logarithm of operator  $B(t)$
with respect to a parameter $t$,
\beq
\pa_{t}\ln B(t)=B^{-1}(t)\pa_{t}B(t),
\eeq
we find the equation
\beq
\nonumber
A^i&=&\widetilde{A}^i+
\Big(:\exp\{-\frac{i}{\hbar}(\widetilde{\Gamma}[\widetilde{{\cal A}}]
\overleftarrow{\pa}_{\!\widetilde{{\cal A}}})h(\widehat{\widetilde{{\cal A}}})\}:
\exp\{{\rm sTr}\ln M(\widehat{\widetilde{{\cal A}}})\}\Big)^{-1}\times\\
\label{iAdA}
&&\quad\times \big(:\exp\{-\frac{i}{\hbar}(\widetilde{\Gamma}[\widetilde{{\cal A}}]
\overleftarrow{\pa}_{\!\widetilde{{\cal A}}})h(\widehat{\widetilde{{\cal A}}})\}:
h^i(\widehat{\widetilde{{\cal A}}})
\exp\{{\rm sTr}\ln M(\widehat{\widetilde{{\cal A}}})\}\big)\cdot 1,
\eeq
allowing to express the fields $A^i$ in terms of $\widetilde{A}^i$.
Using this result in relation (\ref{iGdG}) we obtain the final expression
\beq
\nonumber
&&\Gamma[{\cal A}]=\widetilde{\Gamma}[\widetilde{{\cal A}}]+
\widetilde{\Gamma}[\widetilde{{\cal A}}]\overleftarrow{\pa}_{\!\widetilde{{\cal A}}^i}
\Big(:\exp\{-\frac{i}{\hbar}(\widetilde{\Gamma}[\widetilde{{\cal A}}]
\overleftarrow{\pa}_{\!\widetilde{{\cal A}}})h(\widehat{\widetilde{{\cal A}}})\}:
\exp\{{\rm sTr}\ln M(\widehat{\widetilde{{\cal A}}})\}\Big)^{-1}\times\\
\nonumber
&&\qquad\qquad\qquad\times \big(:\exp\{-\frac{i}{\hbar}(\widetilde{\Gamma}[\widetilde{{\cal A}}]
\overleftarrow{\pa}_{\!\widetilde{{\cal A}}})h(\widehat{\widetilde{{\cal A}}})\}:
h^i(\widehat{\widetilde{{\cal A}}})
\exp\{{\rm sTr}\ln M(\widehat{\widetilde{{\cal A}}})\}\big)\cdot 1-\\
\label{iGdGf}
&&\qquad\qquad\qquad-
i\hbar \ln\big(:\exp\{-\frac{i}{\hbar}(\widetilde{\Gamma}[\widetilde{{\cal A}}]
\overleftarrow{\pa}_{\!\widetilde{{\cal A}}})h(\widehat{\widetilde{{\cal A}}})\}:
\exp\{{\rm sTr}\ln M(\widehat{\widetilde{{\cal A}}})\}\big) \cdot 1,
\eeq
The expression (\ref{iGdGf}) is a general relationship between the effective actions
of two gauge theories obtained one from another by deformation procedure.
Although the relation
(\ref{iGdGf}) looks complicated enough, it is the exact result
in quantum gauge field theory.
Turn
attention to that that this relation is valid for the theories both with closed and open
gauge algebras.

In many applications, it is sufficient to find the effective actions for
the fields satisfying the
equations of motion. Taking into account the possibility of such applications,
one considers the obtained
expression (\ref{iGdGf}) on the equations of motion. The exact equations
of motion for deformed effective
action is written as follows
\beq
\widetilde{\Gamma}[\widetilde{{\cal A}}]\overleftarrow{\pa}_{\!\widetilde{{\cal A}}^i}=0.
\eeq
In this case the relation (\ref{iGdGf}) reduces to the following simple enough form
\beq
\Gamma[{\cal A}]=\widetilde{\Gamma}[\widetilde{A}]-
i\hbar\; {\rm sTr}\ln M(\widehat{\widetilde{{\cal A}}}) \cdot 1
\eeq

Now one finds the expression for deformed effective action through effective
action of initial
theory for the fields on shell. First, find that inverse relations to (\ref{iAdA})
\beq
\label{Atilde}
&&\widetilde{A}^i=A^i+
\Big(:\exp\{-\frac{i}{\hbar}({\Gamma}[{\cal A}]
\overleftarrow{\pa}_{\!\cal A})h(\widehat{{\cal A}})\}:
\exp\{-{\rm sTr}\ln M(\widehat{{\cal A}})\}\Big)^{-1}\times\\
&&
\qquad\times \big(:\exp\{-\frac{i}{\hbar}(\Gamma[{\cal A}]
\overleftarrow{\pa}_{\!{\cal A}})h(\widehat{{\cal A}})\}:
h^i(\widehat{{\cal A}})
\exp\{-{\rm sTr}\ln M(\widehat{{\cal A}})\}\big)\cdot 1.
\eeq
Therefore
\beq
&&\widetilde{\Gamma}[\widetilde{{\cal A}}]=\Gamma[{\cal A}]+
\Gamma[{\cal A}]\overleftarrow{\pa}_{\!{\cal A}^i}
\Big(:\exp\{-\frac{i}{\hbar}(\Gamma[{\cal A}]
\overleftarrow{\pa}_{\!{\cal A}})h(\widehat{{\cal A}})\}:
\exp\{-{\rm sTr}\ln M(\widehat{{\cal A}})\}\Big)^{-1}\times\\
\nonumber
&&\qquad\qquad\qquad\times \big(:\exp\{-\frac{i}{\hbar}(\Gamma[{\cal A}]
\overleftarrow{\pa}_{\!{\cal A}})h(\widehat{{\cal A}})\}:
h^i(\widehat{{\cal A}})
\exp\{-{\rm sTr}\ln M(\widehat{{\cal A}})\}\big)\cdot 1-\\
\label{dGiGf}
&&\qquad\qquad\qquad-
i\hbar \ln\big(:\exp\{-\frac{i}{\hbar}(\Gamma[{\cal A}]
\overleftarrow{\pa}_{\!{\cal A}})h(\widehat{{\cal A}})\}:
\exp\{-{\rm sTr}\ln M(\widehat{{\cal A}})\}\big) \cdot 1,
\eeq
Here the notation
\beq
\widehat{{\cal A}^i}={\cal A}^i+i\hbar
(\Gamma^{(-1)}[{\cal A}])^{ij}
\frac{\overrightarrow{\delta}}{\delta {\cal A}^j}
\eeq
is used and the operators $\widehat{{\cal A}^i}$ commute,
$[\widehat{{\cal A}^i},\widehat{{\cal A}^j}]=0$.
The matrix $(\Gamma^{(-1)}[{\cal A}])^{ij}$
is inverse to the matrix
\beq
\Gamma[{\cal A}]_{ij}=
\frac{\overrightarrow{\delta}}{\delta {\cal A}^i}
\Gamma[{\cal A}]\frac{\overleftarrow{\delta}}{\delta {\cal A}^j},\qquad
\Gamma[{\cal A}]_{ik}
(\Gamma^{(-1)}[{\cal A}])^{kj}=\delta_i^{\;j},
\eeq
with the following symmetry properties
\beq
\label{symi}
(\Gamma^{(-1)}[{\cal A}])^{ij}=-
(\Gamma^{(-1)}[{\cal A}])^{j\;\!i}
(-1)^{(\varepsilon_i+1)(\varepsilon_j+1)},\quad
\Gamma[{\cal A}]_{ij}=-
\Gamma[{\cal A}]_{j\;\!i}
(-1)^{(\varepsilon_i+1)(\varepsilon_j+1)}.
\eeq

On the extremals of the initial effective action, \beq \Gamma[{\cal
A}]\frac{\overleftarrow{\delta}}{\delta {\cal A}^i}=0, \eeq the
equation (\ref{dGiGf}) takes the final simple enough form
\beq
\label{onshell}
\widetilde{\Gamma}[\widetilde{{\cal
A}}]=\Gamma[{\cal A}]+i\hbar\; {\rm sTr}\ln M(\widehat{{\cal A}})\}
\cdot 1
\eeq

Let us briefly discuss the relation of the result (\ref{onshell})
with $S$-matrix structure of the deformed theory. As is known, the
construction of the S-matrix in quantum field theory uses the
effective action calculated on its extremals. If gauge-invariant
deformation of an initial action is performed with the help of a
local function $h(A)$ then the matrix $M(A)$ is local as well. The
second term in the r.h.s. (\ref{onshell}) up to the numerical
constant $i\hbar$ is the exponent  of superdeterminant of a local
function. Such superdeterminant is proportional to $\delta(0)$ which
can be dropped off in the dimensional regularization because of
$\delta(0)=0$. In this case the effective actions coincide,
$\widetilde{\Gamma}[\widetilde{{\cal A}}]=\Gamma[{\cal A}]$. The
same is valid for S-matrices. In general, when the function $h(A)$
involves the derivatives and/or it is non-local, the initial and
deformed theories are not equivalent both at classical and quantum
levels.

As a result, we derived the exact relation between quantum effective actions of the two
classical gauge theories obtained one from another by the deformation preserving the
gauge invariance.

\section{Conclusion}
Let us summarize the results. We have generalized the approach to
constructing the gauge-invariant deformations of a given classical
gauge theory proposed in our previous work \cite{BL-21}. Unlike this
work, we described the gauge-invariant deformation for
the classical theories with open gauge algebra. Deformation of the initial
action both for the theories with closed gauge algebras and for the
theories with open gauge algebras is generated by the transformation
of the fields $A^{i} \rightarrow A^{i}+h^{i}(A)$ with the arbitrary
functions $h^{i}(A)$. If these functions are local and do not
contain the derivatives, the deformed theory is classically
equivalent to the initial theory. However, such functions can
contain the derivatives or to be non-local. Then the deformation yields
a higher derivative or non-local gauge theory. In the work
\cite{BL-21}, we argued that the non-local theories, obtained this
way, can have a closed local sector. As a result, the deformation under
construction can be treated as a procedure to create the new gauge
theories on the base of the given gauge theories. Deformations of
action and gauge transformations are generated by arbitrary functions $h^{i}(A),$
which are completely arbitrary. Explicit
forms of the deformed generators and the deformed gauge algebra are
given by the relations (\ref{gdGen}) and (\ref{dga}). If the initial
gauge algebra is closed, the relation (\ref{dga}) is reduced to
expression for deformed algebra in work \cite{BL-21}.

Considering the two arbitrary gauge theories, where one is obtained
from another by deformation procedure, we have derived a relation
between quantum effective actions for such theories. This relation
is given by the expression (\ref{iGdGf}). Although the obtained
relation looks complicated enough, it is an exact relation. Also, we
proved that the relation (\ref{iGdGf}) is essentially simplified on
the extremals of the effective action. In our previous work
\cite{BL-21}, we constructed some examples of how the non-trivial gauge
theories can be constructed by deformation of the free gauge
theories. Taking into account these results, one can hope that the
relation (\ref{iGdGf}) can be used to relate the quantum aspects of
the trivial and non-trivial gauge theories.

We emphasize that all the results above are based only on the
general properties of gauge theories, and they do not appeal to specific
details of concrete theories. In essence, they represent the exact
relations in the classical and quantum theory of gauge fields.

\section*{Acknowledgments}
\noindent
The work is supported by the Ministry of Education of the Russian Federation, project FEWF-2020-0003.


\begin {thebibliography}{99}
\addtolength{\itemsep}{-8pt}

\bibitem{BL-21}
I.L. Buchbinder, P.M. Lavrov, \textit{On a gauge-invariant
deformation\\ of a classical gauge-invariant theory}, JHEP 06 (2021)
097, {arXiv:2104.11930 [hep-th]}.

\bibitem{BV} I.A. Batalin, G.A. Vilkovisky, \textit{Gauge algebra and
quantization}, Phys. Lett. \textbf{B} 102 (1981) 27- 31.

\bibitem{BV1} I.A. Batalin, G.A. Vilkovisky, \textit{Quantization of gauge
theories with linearly dependent generators}, Phys. Rev. \textbf{D}
28 (1983) 2567-2582.

\bibitem{BV2} I.A. Batalin, G.A. Vilkovisky,
\textit{Closure of the gauge algebra,generalized Lie algebra
equations and Feynman rules}, Nucl. Phys. \textbf{\bf B} 234 (1984) 106.

\bibitem{BLT-15}
I.A. Batalin, P.M. Lavrov, I.V.Tyutin, \textit {Finite anticanonical
transformations in field-antifield formalism}, Eur. Phys. J. {\bf C}
75 (2015) 270, {arXiv:1501.07334 [hep-th]}.

\bibitem{BL-16}
I.A. Batalin, P.M. Lavrov, \textit {Closed description of
arbitrariness in resolving quantum master equation}, Phys. Lett.
{\bf B} 758 (2016) 54-58, {arXiv:1604.01888 [hep-th]}.

\bibitem{BLT-21}
I.A. Batalin, P.M. Lavrov, I.V.Tyutin, \textit {Anticanonical
transformations and Grand Jacobian}, arXiv:2011.06429 [hep-th].






\bibitem{BH}
G. Barnich, M. Henneaux, \textit{Consistent coupling between fields
with gauge freedom and deformation of master equation}, Phys. Lett.
{\bf B} 311 (1993) 123-129, {arXiv:hep-th/9304057}.

\bibitem{H}
M. Henneaux, \textit{Consistent interactions between gauge fields:
The cohomological approach}, Contemp. Math. {\bf 219} (1998) 93-110,
{arXiv:hep-th/9712226}.

\bibitem{D}
A. Danehkar, \textit{On the cohomological derivation of Yang-Mills
theory in the antifield formalism}, Journal of High Energy Physics,
Gravitation and Cosmology {\bf 03} No.02 (2017), Article ID:75808,20
pages.

\bibitem{BaBu}
G. Barnich, N. Boulanger, \textit{A note on local BRST cohomology of
Yang-Mills type theories with Abelian factor}, J. Math. Phys. {\bf
59} (2018) 052302, {arXiv:arXiv:1802.03619 [hep-th]}.

\bibitem{BL}
N. Boulanger, S. Leclercq, \textit{Consistent coupling between
spin-2 and spin-3 massless fields}, JHEP {\bf 11}( 2006) 034,
{arXiv:hep-th/0609221}.

\bibitem{BLS}
N. Boulanger, S. Leclercq, P. Sundel, \textit{On the uniqueness of
minimal coupling in higher-spin gauge theory}, JHEP {\bf 08}, (2008)
056, {arXiv:0805.2764 [hep-th]}.

\bibitem{SS}
M. Sakaguchi, H. Suzuki, \textit{On the interacting higher spin
bosonic gauge fields in BRST-antifield formalism}, Prog. Theor. Exp.
Phys. {\bf 2015}, 00000 (23 pages) {arXiv:2011.02689 [hep-th]}.

\bibitem{Super}
S,J, Gates, M.T. Grisaru, M. Ro\'cek, W. Siegel,
\textit{Superspace}, Benjamin/Cummings, Reading, Mass., 1983,
{arXiv:hep-th/0108200}.

\bibitem{BK}
I.L. Buchbinder, S.M. Kuzenko, \textit{Ideas and Methods of
Supersymmetry and Supergravity}, IOP Publishing, 1998.

\bibitem{Harmonic}
A.S. Galperin, E.A. Ivanov, V.I. Ogievetsky, E.S. Sokatchev,
\textit{Harmonic Superspace}, Cambridge Univ. Press, 2001.

\bibitem{brs1}
C. Becchi, A. Rouet, R. Stora, \textit{The abelian Higgs Kibble
Model, unitarity of the $S$-operator}, Phys. Lett.  {\bf B} 52
(1974) 344.

\bibitem{t}
I.V. Tyutin, \textit{Gauge invariance in field theory and
statistical physics in operator formalism}, Lebedev Institute
preprint  No. 39 (1975), arXiv:0812.0580 [hep-th].


\bibitem{FP}
L.D. Faddeev, V.N, Popov,
{\it Feynman diagrams for the Yang-Mills field},
Phys. Lett. {\bf B} 25 (1967) 29.






\end{thebibliography}
\end{document}